\documentclass[12pt]{article}
\usepackage{epsfig}

\def \sect #1 {\setcounter{equation} 0\section{#1}}
\def \be  {\begin{equation}}
\def \ee  {\end{equation}}
\def \ba  {\begin{eqnarray}}
\def \ea  {\end{eqnarray}}
\def \baa {\begin{eqnarray*}}
\def \eaa {\end{eqnarray*}}
\def \bb  {\begin{thebibliography}}
\def \eb  {\end{thebibliography}}

\def \lab #1 {\label{#1}}

\def \fracs #1#2 {\mbox{\small $\frac{#1}{#2}$}}

\def \bin #1#2 {{\left({#1}\atop{#2}\right)}}
\def \as {\relax\ifmmode\alpha_s\else{$\alpha_s${ }}\fi}

\def \al #1 {\frac {\as({#1})}{\pi} }
\def \ds #1 {\ooalign{$\hfil/\hfil$\crcr$#1$}}

\def \QCD {\mbox{{\tiny QCD}}}

\newcommand \bea{\begin{eqnarray}}
\newcommand \eea{\end{eqnarray}}

\def \O {\Omega}

\def\hepph  #1 {{\tt hep-ph/#1}}

\def \pt {{\rm PT}}

\begin{document}

\begin{flushright}
YITP-SB-04-52
\end{flushright}

\vbox{\vskip .5 in}

\begin{center}
{\Large \bf Resummations, Power Corrections
and\\  Interjet Radiation\footnote{
Submitted to proceedings of the
XXXIV International Symposium on Multiparticle Dynamics. July 26 - August 1,
 2004, Sonoma State University, Sonoma County, California, USA.}}

\vbox{\vskip 0.25 in}

{\large George Sterman}

\vbox{\vskip 0.25 in}

{\it C.N.\ Yang Institute for Theoretical Physics,
Stony Brook University, SUNY\\
Stony Brook, New York 11794-3840, U.S.A.}
\end{center}

\begin{abstract}
Resummation in QCD provides insight into
the evolution of final state jets from short to long 
distances, and of accompanying
interjet radiation.   Applications to event shapes, including
the recently-proposed angularities, 
suggest experimental tests of 
the interrelations between weak- and strong-coupling
dynamics.
\end{abstract}

\bigskip

\noindent
{PACS numbers:  11.15.Bt, 12.38.Cy, 12.38.Lg, 12.38.Qk}

\section{Introduction}
Jets are the imprint on final states of dynamics at short
distances, whether from transitions within the standard
model (Drell-Yan annihilation or QCD scattering, for
example) or from the creation of new degrees of freedom.
An individual jet, however, provides scant clues to
its origin.  Indeed, a jet is simply a collection of co-moving
particles in relative isolation.  The number of 
particles in a given jet, and consequently its total
energy and momentum, are never uniquely defined.
Nevertheless, the distributions of particles and momenta
within and between high energy jets encode information
at all scales, from the largest
momentum transfer or mass, through QCD evolution
to the  scale of the strong coupling, $\Lambda_{\rm QCD}$.
From the point of view of quantum field theory, 
the system passes through a stage of weak coupling
at the  shortest distances on to strong coupling and nonperturbative
dynamics at the longest.  Starting with perturbation theory,
resummations provide one bridge between these regimes.

\section{Resummation: Why?  When? How?}

Only the most inclusive observables depend on a single hard
scale.  More common -- and more interesting -- are functions
of two ordered but perturbative scales, 
$Q \gg Q_1 \gg \Lambda_{\QCD}$.  Varying
the lower scale, $Q_1$ allows us
to move continuously between the ultraviolet and the infrared.
Typically, as $Q_1$ decreases, the perturbative series develops
one or two  logarithmic enhancements in the ratio $Q/Q_1$ for every power
of $\as$.
These enhancements may often be organized, that is resummed, to all orders in
perturbation theory.  Examples include enhancements
that are explicit in the perturbative cross section,
such as those in the transverse momentum distributions ($Q_1=Q_T$)
of vector bosons \cite{kul02}, and enhancements that are implicit
in inclusive distributions, as in threshold resummation
for inclusive Higgs production \cite{cat03}, integrated over $Q_T$.

Perturbative calculations in QCD
depend on the fundamental property of asymptotic freedom.
The most tractable perturbative quantities are one-scale 
cross sections that are  fully infrared safe,
that is, cross sections that can be expressed as series in $\as$
with finite coefficients.   
Infrared safe observables are essentially descriptions of the
flow of energy  \cite{sve95}. The class of infrared safe, single-scale cross sections is
limited to $\rm e^+e^-$ total and jet cross sections, however,
and the latter are single-scale only when the masses of the jets 
are comparable to the total energy $Q$.   When jet masses become small
compared to $Q$, it is necessary to resum, even in
these infrared safe cross sections.  

Resummation of two-scale
logarithms can be derived whenever a
cross section (sometimes amplitude) is a product
or convolution of factors that separate the disparate scales $Q$ and $Q_1$
through the introduction of a third scale, the factorization scale, $\mu\gg \Lambda_{\mathrm QCD}$.
Schematically, a factorizable cross section $\sigma$ that depends on
a very large momentum scale $Q$ and a much softer scale $m$
can be written as
$\sigma(Q,m) =  \omega (Q,\mu)\otimes f(\mu,,m)$.
Here $m$ labels the soft scale(s), typically light
quark masses and $\Lambda_{{\rm QCD}}$,
although it may also represent the lower of two
perturbative scales (say a jet mass) in an infrared safe cross
section.  Whenever there is such a factorization, there is evolution.
Since the physical cross section cannot depend on
the factorization scale, the variation of the short distance function, $\omega$, with $\mu$ must
cancel that of the long distance function, $f$,
\bea
\mu{d\over d\mu} \ln \sigma_{\rm phys}(Q,m) = 0
\quad \Rightarrow \quad  
\mu{d \ln f\over d\mu}= - P(\as(\mu)) = - \mu{d \ln \omega \over d\mu}\, ,
\label{evol}
\eea
where the ``kernel" $P$ can depend only on the variables
that the two functions hold in common.
 Wherever there is evolution there is resummation \cite{Con97},
which is simply the solution to the evolution
equation or equations.
An alternative route to resummed cross sections is
based on coherent branching, which analyzes repeated
gluon emissions \cite{Ell96}.  One may think of each branching
as a ``minifactorization", an incremental step from
short to long distances.

Factorization proofs \cite{Col89} are based on the quantum
incoherence of dynamics at incommensurate length
scales, and of the evolution of systems of particles mutually
receding at nearly the speed of light.
These considerations are reflected in
the structure of factorizable cross sections
near an ``elastic" limit,
with final states that may be characterized
by a single hard scattering and a fixed number 
of jets,   
\bea
\sigma(Q,a+b\rightarrow N_{\rm jets}) 
=  {H} \otimes  {\cal P}_{a'/a}\otimes {\cal P}_{b'/b}
 \otimes {\prod_{i=1}^{N_{\rm jets}}\ J_{i}}\ 
  \otimes{S}\, .
  \label{jetsoft}
\eea
The convolutions may be in partonic momentum fractions,
transverse momenta or energies, depending on the observable.
Physics at the hard scale is in $H$,
while the remaining functions generate
perturbative logarithms and nonperturbative dependence.
In effect, the calculation of any such
cross section can be broken down into a set of standard components:
the product of ${\cal P}$'s, which describe how two partons $a'$ and $b'$, which
result from the quantum evolution of 
partons $a$ and $b$,  collide at $H$, leaving behind forward
jets of  ``spectators", and producing
a set of outgoing jets, $J_{i}$ and coherent soft emission $S$. 
For example,  in hadronic collisions
when partonic energy is just sufficient  
for W or Z production, the cross section takes the form
 $H \otimes {{\cal P}_{q/a}\otimes {\cal P}_{\bar{q}/b}}
  \otimes S$.  Such an inclusive
  cross section is sensitive to long distance
  dynamics only through the forward jets  and the soft radiation.
Similarly, for
$\rm e^+e^-\to 2{\rm J}$ 
the cross section  can be factorized into
a hard part times two jets,
  ${H} \otimes {\ J_{q}\otimes  J_{\bar q}}
   \otimes {S}$,
   while a
  DIS structure function near $x=1$ breaks up into 
   ${H} \otimes {{\cal P}_{q/a}} \otimes {J_{q}}
   \otimes {S}$.  
   Because the cross sections factorize in these
   limits, their logarithmic dependence on
   the masses of the jets and on $1-x$, respectively,
   may be resummed \cite{Con97,Ell96,cat92}.

\section{Application: Angularities in $\rm e^+e^-$}

Among the many applications of resummation, event shapes near the
two-jet limit in $\rm e^+e^-$ annihilation have received perhaps
the most attention, in large part because of the large lever arm
provided by the LEP data, both in the large scale,
the total c.m.\ energy, and in the smaller scale, typically  the jet mass.  

A  generalized class of event shapes, the ``angularities",
were proposed in Ref.\ \cite{ber03a} (reanalyzed and renamed
by Berger and Magnea
in \cite{ber04}), with the motivation of providing a new
parameter that interpolates between distinct traditional measures
of jet substructure.
For final state $N$, we define
\bea
\tau_a={1\over Q}\ \sum_{i\, {\rm in}\, N} 
\, E_i\, \left(\sin\theta_i\right)^a\, \left(1-|\cos\theta_i|\right)^{1-a}
\equiv {1\over Q}\ \sum_{i\, {\rm in}\, N} 
\, E_i\, w_a(\cos\theta_i)\, .
\label{tauadef}
\eea
Here $\theta_i$ is the angle between the direction of
final-state particle $i$ and the thrust axis.  Special cases are $a=0$,
the thrust, $a=1$ broadening, and the total cross section, $a\to -\infty$.
The ``elastic" limit for any finite $a$ is $\tau_a=0$,
where the final state consists of two perfectly collimated jets.
Following the reasoning outlined above to
next-to-leading logarithm (NLL) in $\tau_a$, the
differential cross section factorizes in the space
of Laplace moments  \cite{ber03a},
\bea
\sigma\left(\tau_a,Q,a \right) \!
&=& \sigma_{\rm tot}\; \int_C d\nu\, {\rm e}^{\nu\, \tau_a}\; 
\left[\, { J_{i}(\nu,p_{Ji})}\, \right]^2\, ,
\nonumber\\
{ J_{i}({\nu,p_{Ji})}}
&=&
\int_0 d\tau_a\, {\rm e}^{-\nu\tau_{Ji}}\; 
{ J_{i}({\tau_{Ji},p_{Ji})}}= {\rm e}^{{1\over 2}E(\nu,Q,a)}\, ,
\label{trans}
\eea
into a product of jet functions, $J$.
At the level of NLL, it is possible to
{\it define} $S_{c\bar{c}}=1$, which essentially
serves as a definition of the jets.  This definition
is equivalent to the calculation of the cross section
in terms of jets evolving independently according
to coherent branching \cite{Ell96}.

Logarithms in the  transform variable $\nu$ are
in close correspondence with those in $\tau_a$.
In transform space when $a<1$, logs of $\nu$ exponentiate
in Sudakov form, with up to $n+1$ logarithms
at order $\alpha_s^n$ in the exponent
 $E$ in Eq.\ (\ref{trans}), given in terms of known
 anomalous dimensions $A(\as)$ and $B(\as)$ by
\bea
E(\nu,Q) &=&
  2\, \int\limits_0^1 \frac{d u}{u} \Bigg[ \,
      \int\limits_{u^2 Q^2}^{u Q^2} \frac{d p_T^2}{p_T^2}
A\left(\as(p_T)\right)
      \left( {\rm e}^{- u^{1-a} \nu \left(p_T/Q\right)^{a} }-1 \right)
\nonumber \\
   & & \hspace{5mm} 
     + \frac{1}{2} B\left(\as(\sqrt{u} Q)\right) \left( {\rm e}^{-u
\left(\nu/2\right)^{2/(2-a)} } -1 \right)
      \Bigg]\, . 
\label{Eexpress}
\eea
A characteristic feature of resummed cross sections, illustrated by this expression,
is an integral over scales of the running coupling.  Taken literally,
these expressions are ill-defined, from regions where the integration
variable  $p_T$ is of order $\Lambda_{\mathrm QCD}$.
Although this singularity can be avoided while retaining NLL
accuracy \cite{cat92}, it is also useful to study the
implications of such ambiguities, as we shall see below.

The $a$-dependent expression (\ref{Eexpress}) has yet to be
confronted with real LEP data, except for the case $a=0$,
the thrust. 
Fig.\ 1 shows data for the  closely-related heavy jet
mass distribution at the Z pole \cite{kor00}.
The perturbative-only NLL prediction has the right shape 
overall, but is shifted toward smaller values of jet mass.
We will attribute this shift to nonperturbative corrections below.

\begin{figure}[h]
\centerline{\epsfxsize=8cm \epsffile{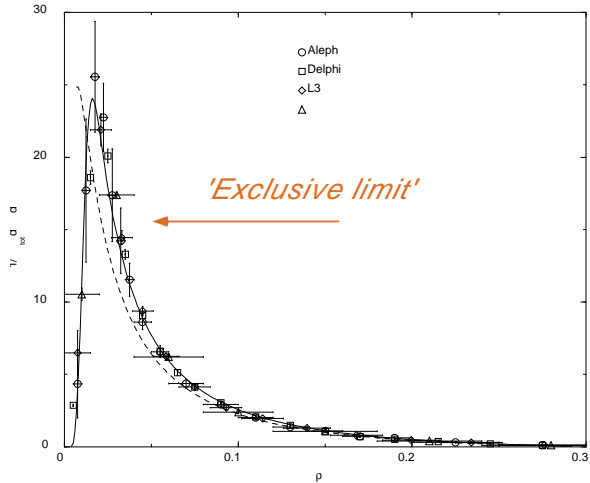}}
\label{heavyfig}
\caption{Heavy jet mass
distribution at the Z \cite{kor00}.  Dashed line: NLL resummed.}
\end{figure}
These considerations may be generalized to 
 jet shapes in deeply inelastic lepton-hadron and  
 hadron-hadron scattering when the  overall final state can be characterized
 by a definite set of jets accompanied by
 soft radiation, as in Eq.\ (\ref{jetsoft}) above.  Such cross sections,
 were termed ``global" by Dasgupta and Salam in Ref.\ \cite{das01}.  
 Recently, Banfi, Salam and Zanderighi have 
extended NLL resummation to a wide class
of global observables in $\rm e^+e^-$, DIS and hadron-hadron scattering,
by developing an innovative software package \cite{ban04}.

\section{Non-global Logs: Color and Energy Flow}

\begin{figure}[h]
\centerline{\epsfxsize=8cm \epsffile{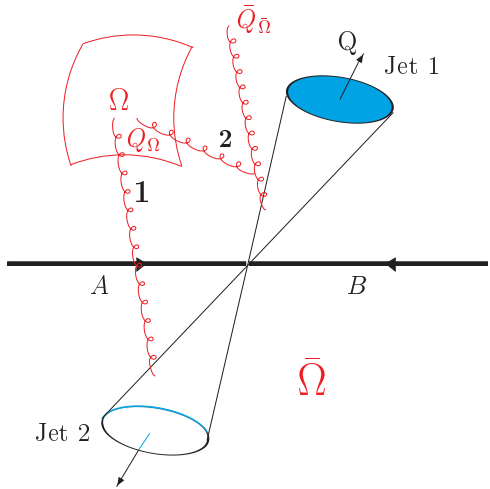}}
\caption{Geometry for energy flow observables.}
\label{flowfig}
\end{figure}

Complementary to jet shapes
are descriptions of interjet energy flow.
  A  simple illustration is shown in Fig.\ \ref{flowfig},
where we trigger on two jet events in the scattering
of particles A and B, and measure the inclusive distribution 
$\Sigma_\Omega (E)$, where $E\ge E_\O\ge 0$, with  
$E_\O$ the energy 
that flows into some angular region $\Omega$, away from
both the collision axis and the jet directions.  Quantities
like $\Sigma_O(E)$ are sometimes referred to as
radiators.

We can imagine (at least) two choices for such a cross section.
First, it may be fully inclusive in the region $\bar{\Omega}$ 
between $\Omega$ and the jets.  In this case,
the number of jets is not fixed, and the observable is
{\it nonglobal} in the terminology Dasgupta and Salam \cite{das01}.
This observable cannot be factorized into a fixed number of
jets as in Eq.\ (\ref{jetsoft}), and as such
cannot be resummed to a simple exponential in the same way as the
event shapes described above.   
Alternatively, we may limit radiation into region $\bar\Omega$
by constructing a correlation with 
an event shape such as $\tau_a$ that
fixes the number of jets  \cite{ber03a,dok03}.

Cross sections where the  number of jets is not fixed are not
fully understood, but they remain infrared safe, so that we
should be able to learn about them in perturbation theory.
Indeed, Banfi, Marchesini and Smye \cite{ban02} showed
that at leading logarithm, $\as^n\ln^n(\sqrt{S}/E_\O)$, 
and in the limit of large numbers of colors, $N_c$,
these cross sections obey a beautiful nonlinear
evolution equation,
\bea
\partial_\Delta\Sigma_{ab}(E)
=-\partial_\Delta R_{ab}\, \Sigma_{ab}(E)
+
\int_{k\ {\rm not\ in}\ \Omega} {d N_{ab\rightarrow k}}\, 
\left(\Sigma_{ak}(E)\Sigma_{kb}(E) - \Sigma_{ab}(E)\right)\, ,
\label{nonlinear}
\eea
where $\partial_{\Delta}=E\partial_E$.
Here $dN_{ab\to k}$ describes the angular distibution of radiation from
a pair (dipole) of color charges, and $R_{ab}$ the corresponding logarithmic
angular integration, where the $\beta$ are four-velocities of partons within
the jets that
radiate soft gluons directly into $\O$ (case 1 in Fig.\ \ref{flowfig}),
\bea
{dN_{ab\to k} = {d\Omega_k\over 4\pi}\;
{\beta_a\cdot\beta_b\over \beta_k\cdot\beta_b\,  \beta_k\cdot\beta_a}}
\quad\quad 
R_{ab} = \int_E^Q {dE'\over E'}\ \int_\Omega {dN_{ab\rightarrow k}}\, .
\label{RNdef}
\eea
These quantities control the linear term on the right-hand side of Eq.\ (\ref{nonlinear}).
The nonlinear terms describe radiation from a hard gluon
of momentum $k$ into region $\bar\O$.  This gluons acts as a new, recoil-less source
of further emission into region $\O$  (2 in Fig.\ \ref{flowfig}).
In the large-$N_c$ limit, any gluon may be thought of as a pair of
sources $q(k)\bar q(k)$, with quark-antiquark color.  In this way, 
for large $N_c$, the combination $\bar{q}(a)G(k) q(b)$ is equivalent to
 an independent pair of dipole sources, $\bar{q}(a)q(k) \oplus \bar{q}(k)q(a)$.
An intriguing relation has been pointed out of this equation with small-$x$ evolution
in the unitarity (saturation) limit  \cite{bal95}.

In the more restrictive
approach, the correlation with an event shape, for example an angularity, 
can fix the number of jets by setting
$\tau_aQ \sim E$ \cite{ber03a}, and we may 
factorize and resum as above,
${d\sigma/dE d\tau_a} \sim 
S(E/\tau_aQ)\otimes {d\sigma_{\rm resum}/d\tau_a}$.
In a combination  of both approaches \cite{dok03},
one may trigger on two narrow
jets, and take the limit $E/(\tau_aQ) \to 0$, using the nonlinear evolution 
above for
the function $S$ that describes radiation into the  interjet region.

With  methods such as these, we can study the
influence of color flow at short distances on energy flow at wide angles
\cite{dok89}, including
applications to final state rapidity gaps \cite{ode99}.

\section{ From Resummed Perturbation Theory to\\ Nonperturbative QCD}

We now return to the interpretation of expressions like
(\ref{Eexpress}), where the argument of $\as$ vanishes in the infrared limit,
As required by infrared safety, if we reexpand
the running coupling in terms of  $\alpha_s(Q)$ for
any fixed momentum scale $Q$, the result is finite at all orders.
The divergence associated with the running coupling comes
from an $n!$ behavior of this expansion at order $\as^n$ \cite{ben03}.   
We treat such contributions as ambiguities in the
perturbative expansion that may be resolved by
supplementing the series with nonperturbative 
parameters, which turn out to be associated with power corrections in the
large scale $Q$.  
Shape functions \cite{kor99} organize the dominant
corrections for event shapes such as angularities to all powers of $Q$.

To be specific, in Eq.\ (\ref{Eexpress}), we treat the region $p_T>\kappa$
using perturbation theory, with $\kappa>\Lambda_{\mathrm QCD}$
a new factorization scale.  For $p_T<\kappa$, we expand the integrand
and replace the complete series of powers
of $\nu p_T/Q$ by $\tilde{f}_{a,\rm NP}$, the shape function,
\bea
E(\nu,Q,a)     &=&   E_\pt
{+\,
\frac{2}{1-a}\
\sum_{n=1}^\infty\ \frac{1}{n\, n!}\, {\left(-{\nu\over Q}\right)}^{n}
\! \int\limits_{0}^{\kappa^2} {dp_T^2\over p_T^2}\; p_T^n\;
A\left(\alpha_s(p_T)\right)}
\equiv E_{\pt}
+ {\ln \tilde f_{a,{\rm NP}}\left(\frac{\nu}{Q},\kappa \right)}\, .
\label{Eexpress2}
\eea
The jet function is now factorized in 
$\nu$-space into perturbative
and nonperturbative functions, resulting in a convolution
after the inverse transform (\ref{trans}),
\bea
\sigma(\tau_a,Q) =  \int_0^{\tau_aQ} d\xi\; f_{a,{\rm NP}}(\xi)\ \sigma_{\rm PT}(\tau_aQ-\xi,Q)\, .
\label{convol}
\eea
Because the momentum space shape function is independent
of $Q$, once it is chosen to describe
the $\tau_a$-distribution  at
some fixed $Q$, say the Z mass, it gives predictions for all $Q$
\cite{ber04,kor99,bel01,ber03b}.

\begin{figure}[h]
\centerline{\epsfxsize=7cm \epsffile{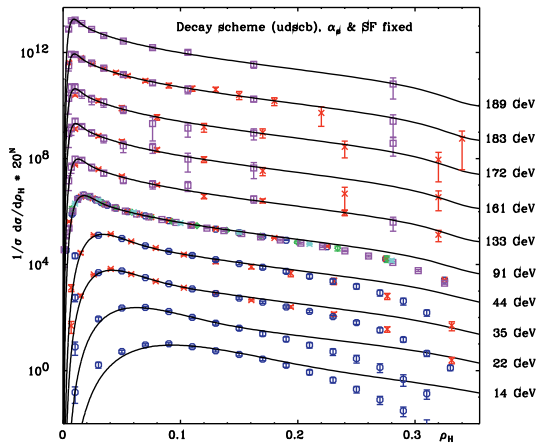}}
\caption{Jet mass fit with a shape function at the Z mass, with
predictions and data for other energies.}
 \label{thrustfig}
\end{figure}

 Fig.\  \ref{thrustfig} illustrates for the heavy jet mass the  quality of fit 
 that may be achieved in this way for a very wide range in $Q$.
What we learn from such an event shape is illustrated by 
the functional form discussed in \cite{bel01}
$f_{0,\rm NP}(\rho) = {\rm const}\ \rho^{c-1}\; {\rm e}^{-d\rho^2}$.
The parameter $c$ may be interpreted in terms of the transverse
momenta emitted per unity rapidity range, while $d$ is related
to  the flow of energy between the hemispheres associated
with the  jets.  More generally, shape functions are related to
correlations of energy flow.  This connection may be made in a manner
that recalls the moment analysis of multiplicity distributions \cite{bia86}.  We
introduce an energy flow operator ${\cal E}(\O)$ by its action on states,
\bea
{\cal E}(\O)|k_1,\dots k_N\rangle = \sum_{j=1}^N k_j^0\delta^2(\O-\O_j)|k_1,\dots k_N\rangle\, .
\label{calEdef}
\eea
For $\tau_a\to 0$, the final state is characterized
by two narrow jets accompanied by soft radiation.
In this limit,  moments of the shape function
may be represented as \cite{kor99}
\bea
\int_0^\infty d\xi\, \xi^n f_a(\xi)
=
\int \prod_{j=1}^n\, d\O_i\; w_a(\cos\theta_j)
\langle {\cal E}(\O_1)\dots {\cal E}(\O_n)\rangle\, ,
\label{calEavg}
\eea
where the angular functions $w_a(\cos\theta)$, have
been defined above in Eq.\ (\ref{tauadef}).
The expectations in (\ref{calEavg}) are taken in the presence
recoilless color sources (Wilson lines) that accurately
represent the coupling of the soft radiation to the jets.
The logarithm of the shape functions is a cumulant expansion
in these expectations \cite{bel01}. 

\begin{figure}[h]
\centerline{\epsfxsize=9cm\epsfig{file=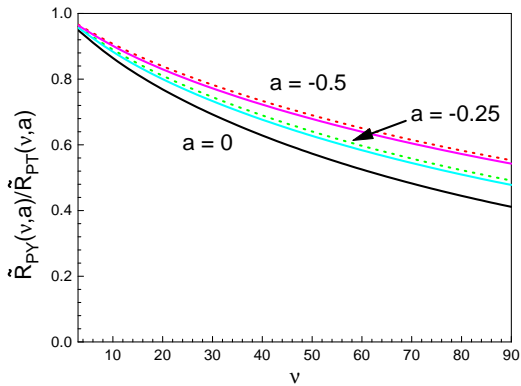,height=7cm,angle=270}}
\caption{Scaling predictions for the shape function in $a$ as tested by PYTHIA.}
\label{tauafig}
\end{figure}

By generalizing the thrust and related
event shapes to angularities, we recognize in Eq.\ (\ref{Eexpress2}) an
interesting scaling property for the associated event  shapes event shapes,
which may be thought of as a test of the rapidity-independence of 
nonperturbative dynamics \cite{ber04,ber03b},
\bea
\tilde f_{a}\left(\frac{\nu}{Q},\kappa\right)
=
\left[\, \tilde f_{0}\left(\frac{\nu}{Q},\kappa\right)\, 
\right]^{1\over 1-a}\, .
\label{scale2}
\eea
It would be interesting to confront this prediciton with LEP data.
For the present, however, we must content ourselves with a comparison
to PYTHIA, which is fairly encouraging, as Fig.\ \ref{tauafig} shows \cite{ber03b}.
It is worth emphasizing that
most event shapes were invented for the $\rm e^+e^-$ jet
physics of the late 70's, and there is more
to learn by addressing  existing data with new analyses
\cite{tka95}.

Also relatively unexplored are
power corrections for single particle inclusive
cross sections.  Here we would like to understand
single-particle cross sections from fixed target to
 collider energies.   
Using  the formalism of Ref.\ \cite{lae00b},
 and the analysis of resummed cross sections 
 described above, it is possible to study the energy
 dependence of power corrections
 in cross sections at measured $x_T=2p_T/\sqrt{S}$ \cite{ste04}.
 For example, in the phase space limit, $x_T\to 1$ for
 $A+B\to \gamma+X$, the cross section may be written as
 the inverse transform of the Mellin moments of the fixed-order cross section
 with a resummation of higher order logarithms, in a form
 analogous to (\ref{Eexpress}),
\bea
p_T^3\,\frac{d\sigma_{AB\to \gamma}}{dp_T}&\sim&
\int_{-i \infty}^{i \infty} \, {dN \over 2\pi i}\;
\tilde{\sigma}_{AB\to \gamma}^{(0)}(N)\;\left(x_T^2\right)^{-N-1}\ {\rm e}^{E_{\mathrm{PT}}(N,p_T)}\;
{\rm e}^{{\delta\,E_{\mathrm{NP}}(N,p_T)}}\; .
\label{1piinverse}
\eea
Isolating  low scales as above, we derive 
a nonperturbative exponent with the $N$ dependence
\bea
\delta\,E_{\mathrm{\rm NP}}(N)= {\rm const.}\,\frac{N^2}{p_T^2}\;\ln\frac{p_T}{N} 
\quad \Rightarrow \quad
\delta\, E_{\mathrm NP}(x_T)=
 {\rm const.}\,\frac{1}{p_T^2\ln^2\left(\frac{4p_T^2}{S}\right)}
\;\ln\left(p_T\ln\left(\frac{4p_T^2}{S}\right)\right)\, ,
\label{Nexp}
\eea
where we have used the conjugate
relation of the variables $N$ and $\ln x_T^2$.
We find a complex behavior in $S$ at fixed $p_T$, 
associated with energy conservation \cite{ste04}.  

\section{Hopeful Conclusion}

Resummations can  bring perturbative QCD to the doorstep of 
nonperturbative field theory.
This analysis is still evolving, and alternative 
treatments of the perturbative/nonperturbative transition
are possible.
Theoretical and experimental studies of  the interplay of color and energy flow 
in hadronic scattering should aid these developments.
Eventually  we will learn to translate fully the language of
partons into the language of hadrons.  

\subsection*{Acknowledgements}

I would like to thank the organizers of the session on jets and of the
International Symposium on Multiparticle Dynamics,
especially Bill Gary, for the  invitation
and encouragement.  I thank my collaborators in some 
of the work reported on here, including Andrei Belitsky, Carola Berger, Gregory Korchemsky, 
Tibor Kucs, Anna Kulesza, Eric Laenen and Werner Vogelsang
for their many insights, and Yuri Dokshitzer, Giuseppe Marchesini,
Gavin Salam and Giulia Zanderighi for discussions. This work was supported in part 
by the National Science Foundation, grants PHY-0098527 and PHY-0354776.

\end{document}